\documentclass[amssymb,aps, prl,superscriptaddress, 10pt]{revtex4-2}

\usepackage{amsmath}
\usepackage{fullpage}
\usepackage{amssymb}
\usepackage{amsfonts}
\usepackage{color}
\usepackage[pdftex]{graphicx}
\usepackage{epstopdf}
\usepackage{soul}
\usepackage{xcolor}
\usepackage{multirow} 
\usepackage{caption}
\usepackage{subcaption}
\usepackage{cases}

\newcommand{\hfds}{HfS$_2$}
\newcommand{\lqd}{\lambda_{q,d}}

\begin{document}

\title{Reflectivity-based refractive index measurement of van der Waals materials}

\author{Xavier Zambrana-Puyalto}
\email{xavislow@protonmail.ch}
\affiliation{Department of Physics, Technical University of Denmark, Fysikvej, DK-2800 Kongens Lyngby, Denmark}
\author{Alexander Johan Olsen}
\affiliation{Department of Physics, Technical University of Denmark, Fysikvej, DK-2800 Kongens Lyngby, Denmark}
\author{Søren Raza}
\email{sraz@dtu.dk}
\affiliation{Department of Physics, Technical University of Denmark, Fysikvej, DK-2800 Kongens Lyngby, Denmark}

\begin{abstract}
We present a reflectivity-based method for measuring the in-plane refractive index of transparent van der Waals (vdW) materials. The approach enables the characterization of as small as $3 \times 3$~{\textmu}m$^2$ exfoliated flakes on a non-transmissive substrate without assuming any specific spectral shape of the refractive index. Exfoliated flakes are most commonly obtained through mechanical exfoliation, which generally produces vdW flakes with tens-of-micron lateral dimensions. As a result, conventional ellipsometry— which depends on large, uniform areas and specific spectral models— becomes challenging to apply. Our method determines the refractive index directly from the spectral position of reflectivity minima, provided the flake thickness and the substrate complex refractive index are known. We demonstrate the technique on hafnium disulfide (HfS$_2$), a vdW semiconductor with high refractive index and low absorption, retrieving its in-plane refractive index across the visible range. The results both validate previous ellipsometry measurements and establish this method as an accessible and spectral-model-free alternative for refractive index characterization of vdW materials. 
\end{abstract}

\maketitle

\section{Introduction}
Van der Waals (vdW) semiconductors have recently emerged as a versatile platform for nanophotonics and optoelectronics~\cite{Zotev2025}. Beyond their well-known excitonic effects in the monolayer limit, some vdW semiconductors in bulk form display unusually high in-plane refractive indices~\cite{Munkhbat2022,Zambrana2025} and strong optical anisotropy~\cite{Ermolaev2021}, making them attractive for applications such as dielectric cavities, metasurfaces, and nonlinear optics~\cite{Zotev2022,Caldwell2019,Nauman2021}. Many other vdW materials have also been predicted to possess outstanding optical properties~\cite{Zambrana2025}, yet these must still be experimentally determined. Of particular importance is the refractive index, which is arguably the most fundamental optical property of a material within linear classical optics.
There are many techniques available for measuring the refractive index, each tailored to a particular class of material. For example, optical fibers are widely used to determine the refractive index of liquids in many different configurations~\cite{Silva2014}, whereas ellipsometry, spectrophotometry and critical angle excitation through a prism are some of the most used techniques to measure the refractive index of solids~\cite{Meyzonnette2019}. 
For vdW semiconductors, one possibility is to use quantitative phase imaging, but this requires transmissive flakes and operates only at a single wavelength at a time~\cite{Khadir2017}. Otherwise, ellipsometry has been the most widely used method~\cite{Yoo2022}, but due to the typical beam diameter of the order 200~{\textmu}m, this method requires large, uniform flakes with sizes of at least $300 \times 300$~{\textmu}m$^2$~\cite{Munkhbat2022}. These sample size requirements are not easily attainable for all vdW materials when using mechanical exfoliation techniques. In addition, ellipsometry typically fits the measurement data to predefined spectral models for the refractive index, such as the Tauc--Lorentz model~\cite{Jellison1996}. Imaging ellipsometry, which focuses the ellipsometer beam with a microscope objective, reduces the sample size requirement to the order of $5 \times 5$~{\textmu}m$^2$ and can retrieve both the in- and out-of-plane refractive indices~\cite{Ermolaev2021,Zotev2023}, but the instrument is expensive and still applies predefined spectral models for the refractive index.
An alternative method for measuring the refractive index directly, without spectral assumptions, and from exfoliated flakes with micron-sized dimensions, would be beneficial for confirming imaging ellipsometry measurements and would make refractive index characterization of vdW semiconductors more accessible.

In this work, we introduce a reflectivity-based method to measure the in-plane refractive index of vdW flakes as small as $3 \times 3$~{\textmu}m$^2$. Our method retrieves the in-plane refractive index through the spectral position of their reflectivity minima, without any spectral assumption. Given the thickness of the flake, we can directly relate the spectral position of the reflectivity minima with the in-plane refractive index at that spectral position~\cite{Sarkar2023arxiv}. The method is based on an analytical model, which has three main assumptions: (i) The vdW flake sits on top of a non-transmissive substrate, (ii) the vdW material has a zero extinction coefficient, and (iii) the illumination is normal to the surface of the flake to be probed. Given these assumptions, an analytical expression can be derived that relates the wavelength of the reflection minima to the in-plane refractive index of the flake, its thickness, and the complex refractive index of the substrate. This expression allows us to find the in-plane refractive index of the vdW flake at the spectral position of the reflection minima provided the thickness of the flake and the complex refractive index of the substrate are known. We prepare several HfS$_2$ flakes with different thicknesses, which produce reflectivity minima across the whole visible spectral range. The thinner flakes display only one or two reflectivity minima in the optical range, thus only yielding one or two data points per flake. In contrast, thicker flakes provide up to six data points. Using the reflectivity-based method, we determine the in-plane refractive index of HfS$_2$ in the wavelength range of 460~nm to 850~nm. Our method of measuring the refractive index can be used for any kind of optical material, but is especially useful for vdW materials that cannot be easily exfoliated or grown uniformly on large areas. Moreover, despite the restrictive assumptions of the theoretical model, this method can practically be used in cases where the illumination is not normal (e.g. with microscope objectives with $\mathrm{NA}<0.5$), and the vdW material in consideration has a low (yet not zero) extinction coefficient.

\section{Theory}
\begin{figure}[ht!]
\centering
\fbox{\includegraphics[width=12 cm]{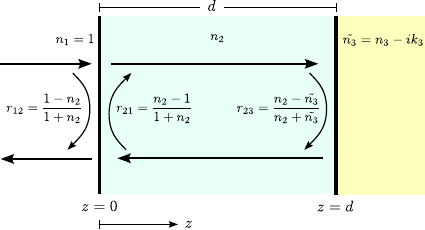}}
\caption{Schematic of a three-layered system, composed of (1) a semi-infinite layer of air, (2) a film of a transparent vdW material of thickness $d$, and (3) a semi-infinite non-transmissive substrate. We denote $r_{kl}$ as the reflection coefficients from medium $k$ to $l$ and $\tilde{n_l} = n_l - i k_l$ as the complex refractive of medium $l$.}
\label{fig:schematics}
\end{figure}
We consider a three-layered system composed of (1) an air superstrate, (2) a vdW thin film, and (3) a non-transmissive substrate (see Fig.~\ref{fig:schematics}). Using the $e^{i(\omega t - \mathbf{k} \cdot \mathbf{r})}$ sign convention and the wave-transfer matrix formalism \cite[p.258,]{Saleh1991}, we can write the reflection coefficient $r$ of the whole system at normal incidence as 
\begin{equation}
r=\dfrac{r_{12}+r_{23}\exp\left[-2 i \beta_2 d\right]}{1+r_{12}r_{23} \exp\left[-2 i \beta_2 d\right]},
\label{eq:r13}
\end{equation}
where $r_{kl}=\frac{\tilde{n_k}-\tilde{n_l}}{\tilde{n_k}+\tilde{n_l}}$ are the reflection coefficients from medium $k$ to $l$ at normal incidence, $\tilde{n_l}=n_l-ik_l$ is the complex refractive index of medium $l$, with $n_l$ and $k_l$ being respectively the refractive index and the extinction coefficient, and $\beta_l=\frac{2\pi n_l}{\lambda}$ is the wavenumber in medium $l$. In general, the reflection coefficients are complex numbers, allowing us to write them in polar form as $r_{kl}=\rho_{kl}e^{-i\alpha_{kl}}$. We only do this for $r_{23}$, as $r_{12}$ is a real number, since we assume that the vdW material does not absorb any light. This allows us to rewrite Eq.~(\ref{eq:r13}) as
\begin{equation}
r=\dfrac{r_{12}+\rho_{23}\exp\left[- i \chi \right]}{1+r_{12}\rho_{23} \exp\left[-i \chi \right]},
\label{eq:r13X}
\end{equation}
with $\chi=2\beta_2 d + \alpha_{23}$. The reflectance of the three-layer system $R=\vert r\vert^2$ is then given by
\begin{equation}
R=\dfrac{r_{12}^2+\rho_{23}^2+2r_{12}\rho_{23}\cos\chi}{1+ r_{12}^2 \rho_{23}^2+2r_{12}\rho_{23}\cos\chi}.
\label{eq:R}
\end{equation} 
The spectral minima and maxima of reflectance are given by the condition $\frac{\partial R}{\partial \lambda}=0$, which yields the solution $\chi = q \pi$, with $q$ being an integer that tracks the reflectivity minima order. The minima are such that $ \frac{\partial^2 R}{\partial \lambda^2 } > 0 $. This condition discards half of the solution set, giving $\chi = (2q)\pi$ if $r_{12}<0$ and $\chi = (2q+1)\pi$ if $r_{12}>0$. As a result, each reflectance minima gives us a relation between all the different variables of the problem. In particular, we can isolate the film thickness $d$ as \cite{Park1964,Lunacek2009}
\begin{equation}
d = \dfrac{\lambda_q}{4\pi n_2} \left[ 2q\pi - \arctan\left( \dfrac{-2n_2k_3}{n_2^2-n_3^2-k_3^2} \right) \right],
\label{eq:main}
\end{equation}
where it has been used that $r_{12}<0$, as the first medium is air. Notice that the refractive indices $n_2$ and $n_3$ and the extinction coefficient $k_3$ are evaluated only at the spectral minima $\lambda_q$. In the next section, we apply Eq.~(\ref{eq:main}) to determine the in-plane refractive index of \hfds. 

\section{Experimental results}
\begin{figure}[ht!]
\centering
\includegraphics[width=12 cm]{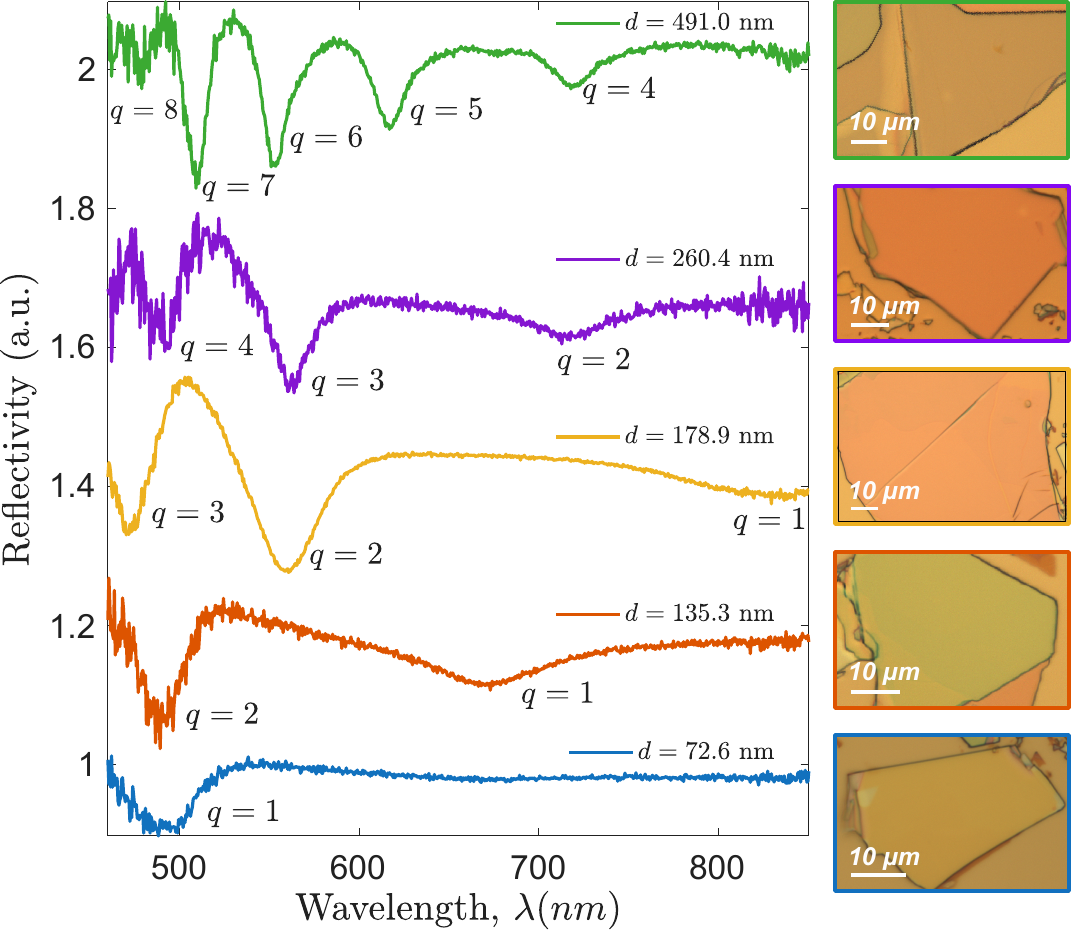}
\caption{Normalized reflectivity spectra of \hfds\ flakes of different thicknesses $d$ on top of a $100$~nm gold layer. For clarity, the curves are offset with $+0.2,+0.4,+0.7,+1.05$ a.u., respectively. The thinnest flake ($d=72.6$~nm) yields only a $q=1$ reflectance dip. The $q=1$ reflectivity minima is observable in our spectral range up to the flake thickness of $d=178.9$~nm. The thickest flake ($d=491.0$~nm) displays five different minima of orders $q=4,..,8$. All curves are normalized to the reflection of the gold substrate. An optical image of each flake is shown to the right of each reflectivity curve.}
\label{fig:reflectivity}
\end{figure}

\hfds\ is a semiconductor material whose optical properties have recently been examined~\cite{Zambrana2025}. It is characterized by a high refractive index in the visible range and a large band gap, which results in a negligible extinction coefficient for $\lambda>500$~nm. We exfoliate \hfds\ onto a $100$~nm thick gold layer, on top of a microscope coverslip. The gold thickness is sufficient to ensure negligible transmission. With the aim of measuring the in-plane refractive index of \hfds, we measure the reflectivity of 17 different \hfds\ flakes, with thicknesses $d$ in the range of 70~nm to 600~nm. The thicknesses of the flakes are measured with a Dimension Icon-PT atomic force mircroscope (AFM) from Bruker AXS in tapping mode. The reflectivity measurements are carried out with an Andor Kymera 328i spectrograph, which measures the light reflected from the flakes under an OSL-2 fiber-coupled unpolarized halogen light source. All the measurements are performed using a $10\times$ microscope objective with a $\textrm{NA}=0.3$ on a Nikon Eclipse LV100ND microscope. We use a $10\times$ microscope objective so that the illumination is as close as possible to the normal incidence conditions of our theoretical model (see Fig. \ref{fig:schematics}). However, we have tested that a $20\times$ microscope objective with a $\textrm{NA}=0.45$ also yields good results and allows us to measure flakes with an area of the order of $10 \ \mu$m$^2$. We normalize the reflectivity of the three-layer system (air-\hfds -gold) by the reflectivity of air-gold, so that the spectral properties of our equipment do not influence the results. The dark counts of the spectrograph are subtracted from both reflectivity measurements. 

\begin{figure}[ht!]
\centering
\includegraphics[width=8.5 cm]{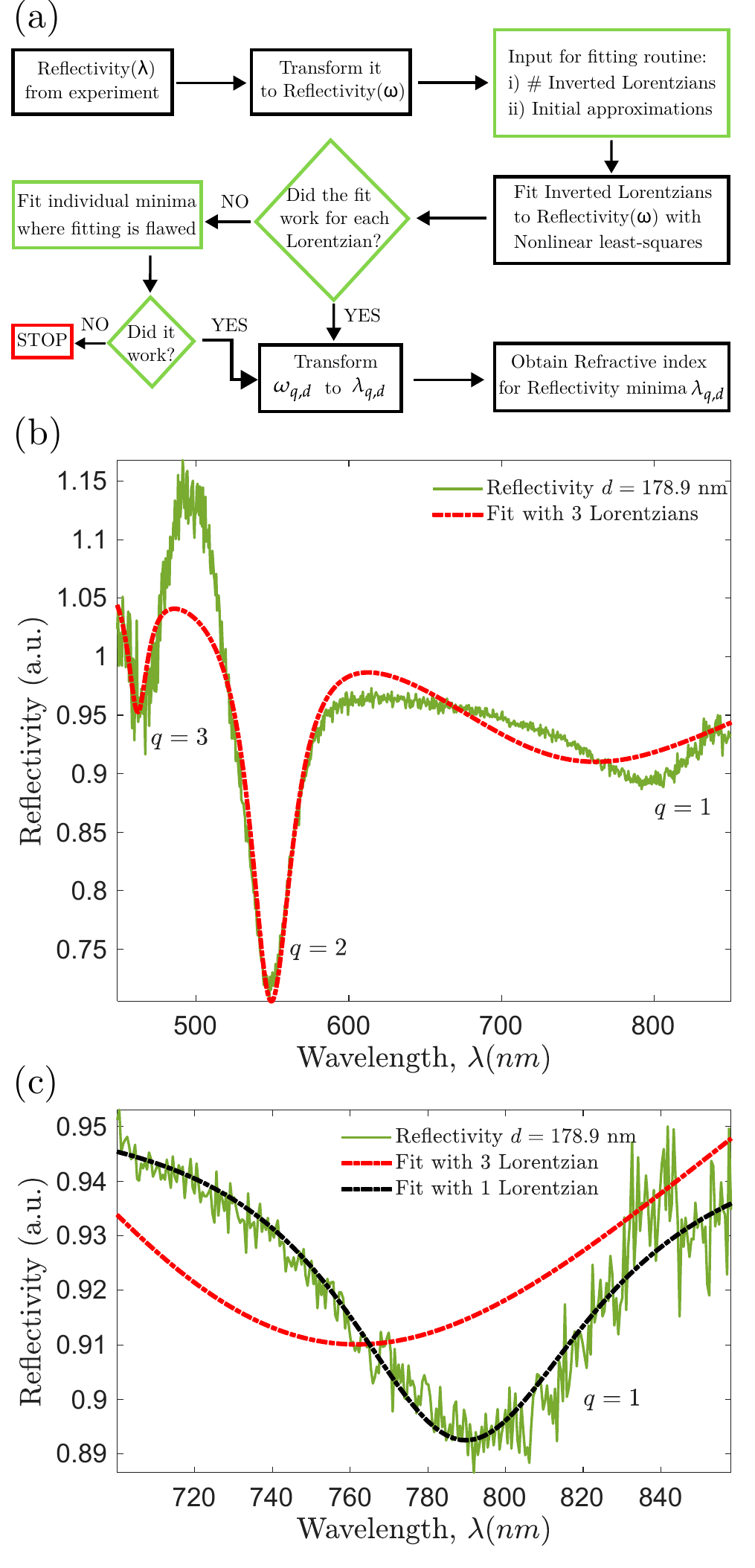}
\caption{(a) Diagram showing the method used to extract $\lqd$ and the refractive index $n_2(\lqd)$ from reflectivity measurements such as the ones shown in Fig.~\ref{fig:reflectivity}. (b) Reflectivity of a \hfds\ flake with a thickness of $d=178.9$ nm along with a 3-Lorentzian fit (see Eq.~(\ref{eq:Lorentzians})). (c) Zoom-in reflectivity plot of the $q=1$ minimum shown in (b), as well as two different fits using a single and a 3-Lorentzian model.}
\label{fig:Lorentzians}
\end{figure}
In Fig.~\ref{fig:reflectivity}, we show the normalized reflectivity of five selected flakes, sorted by their thickness. Except for the reflectivity curve for the thinnest flake, an offset is added to each respective curve. Optical images of each of the measured flakes are also presented (right side of Fig.~\ref{fig:reflectivity}). In the reflectivity measurements, we have marked the reflectivity minima order $q$, as it is one of the required parameters in Eq.~(\ref{eq:main}) to relate the reflection minima to the refractive index. The thinnest flake ($d=72.6 \pm 1.6$~nm) has a reflectivity minimum of order $q=1$ at a wavelength of around 500~nm. The minimum redshifts as the thickness of the flake increases. For the flake with $d=178.9 \pm 0.8$~nm, the same $q=1$ minimum is present at a wavelength of 795~nm, and it also supports two other minima of orders $q=2,3$ at wavelengths around 547 and 465~nm, respectively. The thickest flake ($d=491.0 \pm 1.8$~nm) displays 5 minima of orders $q=4,..,8$. Each reflection minima is associated to a certain thickness of the flake $d$ and to a certain reflection order $q$. Equation~(\ref{eq:main}) can be numerically solved to uniquely relate the minima wavelengths $\lambda_{q,d}$ with the unknown in-plane refractive index of \hfds\ at that wavelength $n_2( \lambda_{q,d} )$. Our optical system displays increasing noise for wavelengths below 500~nm and above 800~nm, especially for lower wavelengths. Nonetheless, we have considered reflectivity data in the wavelength range of 460~nm to 850~nm, as we have observed that our measurements yielded small errors in this extended domain.

We now comment on the method used to extract the refractive index using the reflectivity curves shown in Fig.~\ref{fig:reflectivity} as input. Equation~(\ref{eq:main}) allows for the computation of the refractive index of \hfds\ provided the following parameters are known: the thickness of the flake $d$, the reflectivity minima order $q$, the wavelength of reflectivity minima $\lqd$, and the complex refractive index of the substrate at the wavelength of the reflectivity minima $\tilde{n}_3(\lqd)$. Except for $q$, all the other parameters are prone to some uncertainties. For example, the uncertainty on $d$ is determined by the AFM measurements and is typically below 1~nm. The complex refractive index of the gold substrate could deviate from its tabulated value, which in our case is obtained from Ref.~\cite{Yakubovsky2017}. However, the parameter with the largest uncertainty is the resonance wavelength $\lqd$, because the reflectivity data around the minima is noisy. We can therefore not determine $\lqd$ by simple visual inspection. To extract $\lqd$ consistently, we fit the data to a superposition of inverted Lorentzian curves in frequency space. In our fit, apart from the amplitude $A_i$, the central frequency $\omega_{0,i}$ and the full-widths at half maximum $\gamma_i$ of each Lorentzian, we also include a baseline parameter $B$, which we impose to be in the $0.9-1.1$ range. The multiple-Lorentzian equation that we fit to the measured reflectivity data is therefore given as
\begin{equation}
R\left( \omega \right) = B - \sum_i \dfrac{A_i}{1+ \left( \dfrac{\omega - \omega_{0,i}}{\gamma_i} \right) }.
\label{eq:Lorentzians}
\end{equation}
Our fitting method to determining the resonance wavelength $\lqd$ is visualized in Fig.~\ref{fig:Lorentzians}(a). The first step is to transform our wavelength-dependent reflectivity data to the frequency domain. Subsequently, we fit as many inverted Lorentzians as we can identify. For instance, in the reflectivity curves shown in Fig.~\ref{fig:reflectivity}, we have fitted, from top to bottom, 5, 3, 3, 2, and 1 Lorentzians, respectively. Each Lorentzian yields a central frequency $\omega_{0,i}$, which we transform into a central wavelength $\lambda_{0,i} = 2\pi c /\omega_{0,i}$. After identifying the minimum order $q$ and the thickness of the flake $d$, we directly assign each $\lambda_{0,i}$ to a wavelength minimum $\lqd$. 
In most cases, our script outputs a fit which closely matches the experimental data. However, in some cases, the fit does not manage to describe all the minima. In Fig.~\ref{fig:Lorentzians}(b), we show one of these cases, in particular the \hfds\ flake with $d=178.9 \pm 0.8$~nm (yellow curve in Fig.~\ref{fig:reflectivity}), where only two out of the three minima are properly fitted by the inverted Lorentzian model. In such cases, we proceed to fit a single Lorentzian to each of the reflectivity minima where the multiple-Lorentzian routine failed. In Fig.~\ref{fig:Lorentzians}(c), we show the $q=1$ minima fitted with both a single and 3-Lorentzian model. We observe that the single Lorentzian model allows us to retrieve another data point for $\lqd$ and consequently for $n_2(\lqd)$. The cases in which a central frequency $\omega_{0,i}$ could not be identified using either the multiple- or single-Lorentzian model were excluded from consideration. It is worth noting that our fitting routine also provides the uncertainty associated to the reflectivity minima, i.e., $\delta_{\lqd}$.

\begin{figure}[ht!]
\centering
\includegraphics[width=11 cm]{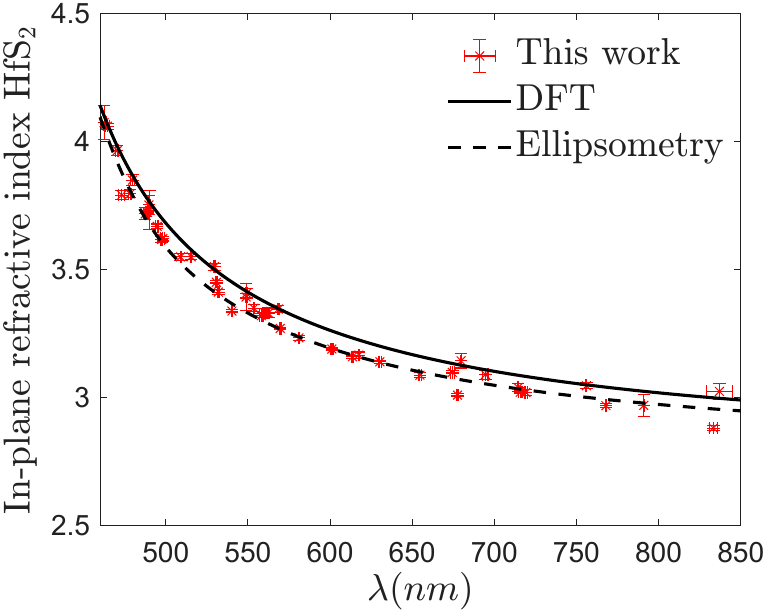}
\caption{In-plane refractive index of \hfds\ measured with the reflectivity-based method described in this work (red dots) compared to density functional theory (DFT) calculations (black line) and imaging ellipsometry measurements (black dashed lines). The spectral range covers wavelengths from 460~nm to 850~nm. The DFT and imaging ellipsometry data are taken from Ref.~\cite{Zambrana2025}.}
\label{fig:RI}
\end{figure}

Following this methodology, we set out to measure the in-plane refractive index of \hfds. We have collected data from 17 different flakes, which provide 49 different spectral minima $\lqd$ in the $\left[ 460,850\right]$~nm spectral range. We have disregarded 4 $\lqd$ values, as neither the multi- nor the single-Lorentzian fit worked properly. By applying Eq.~(\ref{eq:main}), we use the remaining data points for $\lqd$ to obtain 45 unique values for the in-plane refractive index of \hfds. The uncertainty of the refractive index has been computed as $\delta_{n_2} = \sqrt{\left( \dfrac{\partial n_2}{\partial d} \delta_d\right)^2 + \left( \dfrac{\partial n_2}{\partial \lqd} \delta_{\lqd}\right)^2}$, where we have assumed that they are independent variables. The experimental data for the in-plane refractive index of HfS$_2$ is presented in Fig.~\ref{fig:RI}. Our measurements are compared to ab initio calculations and experimental data obtained using imaging ellipsometry, both of which are available in Ref.~\cite{Zambrana2025}. We observe that our measurements closely match both sets of data, in fact they mostly lie in between the two sets, thus confirming the reliability of this method. All data collected in this work are available in an open-access public repository~\cite{Zambrana2025DataRI}.  

\section{Conclusion}
To conclude, we have shown that the in-plane refractive index of a transparent vdW material such as \hfds\ can be measured in the visible range with a reflectivity-based method. The method works for transparent and weakly absorbing materials and can be applied to flakes with lateral dimensions down to the order of $10 \ \mu$m$^2$. We have tested the method with \hfds\ flakes and have been able to retrieve its refractive index. Our results match the theoretical and experimental results shown in Ref.~\cite{Zambrana2025}. Due to its simplicity and accuracy, we believe that the reflectivity-based method presented in this work can make refractive index measurements more accessible as well as provide a means to independently confirm imaging ellipsometry measurements.

\section{Acknowledgments}
X.Z.-P. and S.R. acknowledge funding from the Villum Foundation (VIL50376) and the Novo Nordisk Foundation (NNF24OC0096142).

%

\end{document}